\documentclass[%
 reprint,
superscriptaddress,
 amsmath,amssymb,
 aps,
prl,
]{revtex4-2}

\usepackage{graphicx}
\usepackage{dcolumn}
\usepackage{bm}
\usepackage{xcolor} 

\begin{document}

\preprint{APS/123-QED}

\title{Ultrafast modification of the electronic structure of a correlated insulator}%


\author{O. Grånäs}
\affiliation{
 Department of Physics and Astronomy, Uppsala University, Uppsala, Sweden
}
\email{oscar.granas@physics.uu.se}

\author{I. Vaskivskyi}%
\affiliation{
 Department of Physics and Astronomy, Uppsala University, Uppsala, Sweden
}
\affiliation{Center for Memory and Recording Research, University of California San Diego, 9500 Gilman Drive, La Jolla, CA 92093-0401, USA}

\author{X. Wang}
\affiliation{
 Department of Physics and Astronomy, Uppsala University, Uppsala, Sweden
}

\author{P. Thunström}
 \affiliation{
 Department of Physics and Astronomy, Uppsala University, Uppsala, Sweden
}

\author{S. Ghimire}
\affiliation{Stanford PULSE Institute, SLAC National Accelerator Laboratory, 2575 Sand Hill Road, Menlo Park, CA 94025, USA
}

\author{R. Knut}
\affiliation{
 Department of Physics and Astronomy, Uppsala University, Uppsala, Sweden
}

\author{J. Söderström}
\affiliation{
 Department of Physics and Astronomy, Uppsala University, Uppsala, Sweden
}
\author{L. Kjellsson}
\affiliation{
 Department of Physics and Astronomy, Uppsala University, Uppsala, Sweden
}
\author{D. Turenne}
\affiliation{
 Department of Physics and Astronomy, Uppsala University, Uppsala, Sweden
}
\author{R. Y. Engel}
\affiliation{Department of Photon Science, DESY, Notkestraße 85, D-22607 Hamburg, Germany}
\author{M. Beye}
\affiliation{Department of Photon Science, DESY, Notkestraße 85, D-22607 Hamburg, Germany}
\author{J. Lu}
\affiliation{Stanford PULSE Institute, SLAC National Accelerator Laboratory, 2575 Sand Hill Road, Menlo Park, CA 94025, USA
}
\author{D. J. Higley}
\affiliation{SLAC National Accelerator Laboratory, 2575 Sand Hill Road, Menlo Park, CA 94025, USA}
\author{A. H. Reid}
\affiliation{SLAC National Accelerator Laboratory, 2575 Sand Hill Road, Menlo Park, CA 94025, USA}
\author{W. Schlotter}
\affiliation{SLAC National Accelerator Laboratory, 2575 Sand Hill Road, Menlo Park, CA 94025, USA}
\author{G. Coslovich}
\affiliation{SLAC National Accelerator Laboratory, 2575 Sand Hill Road, Menlo Park, CA 94025, USA}
\author{M. Hoffmann}
\affiliation{SLAC National Accelerator Laboratory, 2575 Sand Hill Road, Menlo Park, CA 94025, USA}
\author{G. Kolesov}
\affiliation{John Paulson School of Engineering and Applied Sciences, Harvard University, Cambridge, Massachusetts 02138, USA }
\author{C. Schüßler-Langeheine}
\affiliation{Helmholtz-Zentrum Berlin für Materialien und Energie GmbH, 12489 Berlin, Germany}
\author{A. Styervoyedov}
\affiliation{Max-Planck Institut für Mikrostrukturphysik, Weinberg 2, Halle, Germany }
\author{N. Tancogne-Dejean}
\affiliation{Max Planck Institute for the Structure and Dynamics of Matter and Center for Free-Electron Laser Science, Luruper Chaussee 149, 22761 Hamburg, Germany}
\author{M. A. Sentef}
\affiliation{Max Planck Institute for the Structure and Dynamics of Matter and Center for Free-Electron Laser Science, Luruper Chaussee 149, 22761 Hamburg, Germany}
\author{D. A. Reis}
\affiliation{Stanford PULSE Institute, SLAC National Accelerator Laboratory, 2575 Sand Hill Road, Menlo Park, CA 94025, USA
}
\author{A. Rubio}
\affiliation{Max Planck Institute for the Structure and Dynamics of Matter and Center for Free-Electron Laser Science, Luruper Chaussee 149, 22761 Hamburg, Germany}
\affiliation{Center for Computational Quantum Physics, Flatiron Institute, New York, NY 10010 USA}
\author{S. S. P. Parkin}
\affiliation{Max-Planck Institut für Mikrostrukturphysik, Weinberg 2, Halle, Germany }
\author{O. Karis}
\affiliation{
 Department of Physics and Astronomy, Uppsala University, Uppsala, Sweden
}
\author{J.-E. Rubensson}
\affiliation{
 Department of Physics and Astronomy, Uppsala University, Uppsala, Sweden
}
\author{O. Eriksson}
\affiliation{
 Department of Physics and Astronomy, Uppsala University, Uppsala, Sweden
}
\affiliation{School of Science and Technology, Örebro University, SE-701 82, Örebro, Sweden}
\author{H. A. Dürr}
\affiliation{
 Department of Physics and Astronomy, Uppsala University, Uppsala, Sweden
}

\date{\today}

\begin{abstract}
A non-trivial balance between Coulomb repulsion and kinematic effects determines the electronic structure of correlated electron materials. The use electromagnetic fields strong enough to rival these native microscopic interactions allows us to study the electronic response as well as the timescales and energies involved in using quantum effects for possible applications. We use element-specific transient x-ray absorption spectroscopy and high-harmonic generation to measure the response to ultrashort off-resonant optical fields in the prototypical correlated electron insulator NiO. Surprisingly, fields of up to 0.22 V/Å leads to no detectable changes on the correlated Ni 3d-orbitals contrary to previous predictions. A transient directional charge transfer is uncovered, a behavior that is captured by first-principles theory. Our results highlight the importance of retardation effects in electronic screening, and pinpoints a key challenge in functionalizing correlated materials for ultrafast device operation.

\end{abstract}


\maketitle




Strongly correlated electron materials underlie many of the most intriguing macroscopic manifestations of quantum physics \cite{Li:2013bu, Lee:2006de}, enabling a multitude of technologically relevant phenomena \cite{Zhang:2014dq}. In transition metal oxides, the inherent competition between the local Coulomb repulsion between the transition-metal $d$-electrons and orbital hybridization often gives rise to a strong correlated electronic structure \cite{Imada:1998ti}. The Coulomb repulsion favors localization of electrons, whereas the orbital hybridization promotes band-formation and allows electrons to hop, from site to site. These two quantities are commonly represented by the Hubbard $U$ and the hopping parameter $t$ respectively.
In correlated electron materials, the response to strong off-resonant electromagnetic fields is highly non-trivial. The induced polarization, and the displacement-field within the material disturb the charge distribution. Sufficiently strong fields may upset the balance between the hopping energy scale $t$ and the Hubbard $U$, potentially driving a material from an insulating to a metallic behavior.

In fact, the prototypical correlated insulator nickel oxide is predicted to exhibit a dramatic change of Hubbard $U$ in response to non-resonant ultra strong fields \cite{TancogneDejean:2018je}. 
The spatiotemporal details of the response of the electronic structure is still an outstanding problem, and this prediction remains to be corroborated by experimental studies. Hence, time-resolved experimental evidence, in particularly if a site- and orbital resolved response can be measured, is of outmost importance. 
With the advent of experimental pump-probe techniques, the response of targeted degrees of freedom can be measured. Still, the ability to probe the response of both localized electrons and those with itinerant character on ultrafast timescales requires a combination of techniques, in particular in the presence of non-resonant strong electromagnetic fields.

NiO exhibits signatures of localized atomic-like states originating with the Ni 3d-orbitals \cite{Groot:2005ew} as well as band-like dispersive electronic states \cite{Shen:1991he}. The localization of Ni 3d-orbitals is a result of a large Hubbard $U$ in relation to the hopping $t$. This balance originates with a weak screening of the local Coulomb-interaction between 3d-electrons by surrounding itinerant electrons. 
The O 2p-states have a fundamental impact on optical and electrical transport properties \cite{Sawatzky:1984jt}, which is captured by the charge-transfer energy-scale, as the energetic overlap allows electron transfer from O 2p-states to the localized Ni 3d-states. This mechanism is also relevant for the super-exchange mechanism responsible for the antiferromagnetic order between Ni-atoms along the Ni-O-Ni bond direction \cite{Betto:2017fh}. The balance between the charge transfer energy scale, the Hubbard $U$ and the hopping $t$ results in a band-gap of about 4 eV in the ground state.

We use off-resonant optical fields to transiently disturb the balance between the governing energy-scales. The selected 0.6 eV photon-energy is midway between the well-known two-magnon resonance at ~0.25eV and the onset of d-d excitations above ~0.8eV \cite{Newman:1959ce}. The narrow bandwidth (100meV FWHM) of the pump pulses implies that no resonant transitions are possible.
We analyze the response through complementary experimental techniques, probing the locality of the electronic response with resonant x-ray absorption spectroscopy (XAS), while investigating the response of hybridized valence states by monitoring the emission of higher harmonics of the optical driving field. Aided by non-equilibrium first-principles simulations we analyze the temporal response of the electronic structure of this correlated system. This allows for an improved understanding of the microscopic interactions of quantum phases, as well as shining light on the timescales and energies involved in using quantum effects for possible applications.

The optical field-strength was varied between 0.11 and 0.22 V/Å corresponding to optical pump fluences of 30 and 120 mJ/cm$^2$, respectively. In this regime multi-photon transitions as well as Zener tunnelling between valence and conduction states are suppressed (see supplemental material II.A). The response of the electronic structure to external electromagnetic fields in this regime is governed primarily by two effects. A modified band-structure according to the dynamical Franz-Keldysh effect \cite{vonFranz:1958ux, Keldysh:1958vu, Novelli:2013gk}, where tails from the valence and conduction states leak into the band gap region. The altered wave-function overlap is expected to induce a dynamic modification of the hopping ($t\rightarrow \tilde{t}$). The modification of the electronic structure can also lead to a dynamic renormalization of the Hubbard $U$. The interaction between localized Ni 3d and itinerant electrons of mainly O 2p electrons screens the Hubbard U from an atomic value of 20-25 eV down to 7-8 eV \cite{Panda:2017ga}. In this scenario, strong electromagnetic fields influence crystal field levels and ligand states, causing significant changes in screening, i.e. a renormalization of the Hubbard U ($U\rightarrow \tilde{U}$). Note that electronic states not hybridizing with the local 3d electrons can affect the Hubbard $U$ through a modified ability to screen the local interactions. A schematic illustration of of these scenarios, can be found in the supplementary material II.B, Fig. S3.

It is difficult to predict which of the two responses dominate for a given material. From a theoretical perspective, recent developments in out-of-equilibrium many-body methods are still at the stage of few-band models \cite{Golez:2017ix}, while full band-structure calculations of out-of-equilibrium dynamics suffer from difficulties including the effect of strong correlation. Here, one resorts either to a time-independent effective Hubbard $U$, or an instantaneous change, representing the two extreme cases of a time-independent or an instantaneous effect on screening \cite{TancogneDejean:2018je}. Predictions based on these different techniques differ qualitatively. An instantaneous screening indicates a modification of $U$ by about 10\% for optical fields up to about 0.2 V/Å \cite{TancogneDejean:2018je}. The influence of the pump with the above mentioned characteristics, is equivalent to a modification of the energy-landscape in the order of 30\% of the bandwidth for electronic states extending over the lattice parameter in the material, or 7500\% of the magnetic super-exchange coupling.  
Using the fact that core-states are well localized at an atomic site, pump-probe XAS allows us to disentangle the response of the hybridized bands to give time- and spatially resolved information of the pump response of this correlated insulator.
Through the knowledge of the band-structure, one may relate the response to modulations of variables in microscopic models based on hopping $t$, and screened Coulomb repulsion $U$. In NiO, transitions for O from 1s core to 2p conduction states and Ni 2p to 3d states are accessible with soft x-rays at photon energies around 535 eV and 853 eV, respectively. Note that XAS does not necessarily probe the dynamics of the conduction band edge, but the dynamics of the final-state orbitals of the absorption process.

\begin{figure}
     \includegraphics[width=\linewidth]{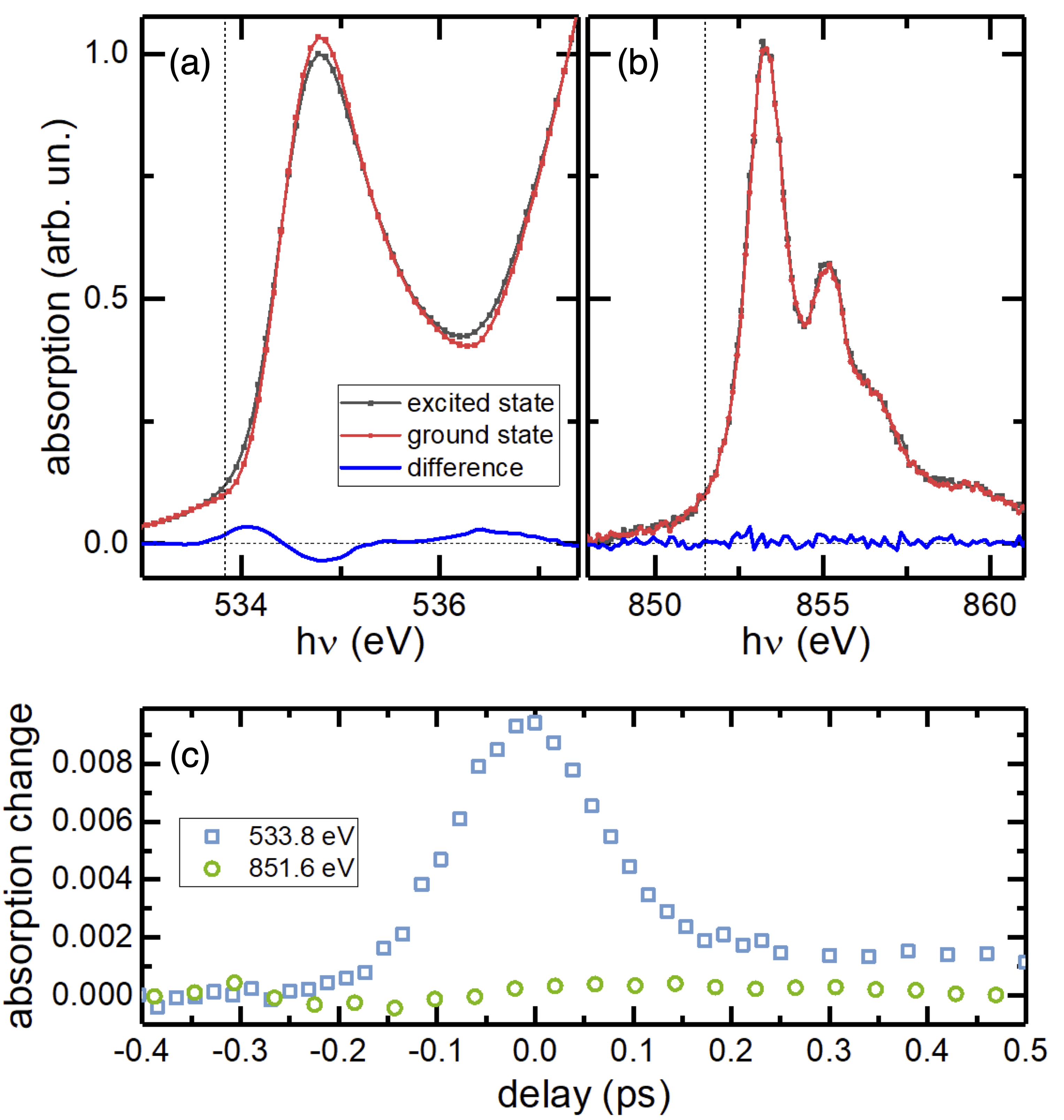}
    \caption{(a) Measured XAS spectra for O 1s-2p transitions and (b) for Ni 2p-3d transitions. Shown are measurements with (black lines) and without (red lines) the driving laser, as well as the difference (blue lines) for the laser electric field E parallel to [100] with simultaneous arrival time of laser pump and x-ray probe. (c) Temporal evolution of the changes in O (blue squares) and Ni (green circles) XAS vs. pump-probe time-delay. The peak around zero time-delay of the O XAS represents the response shown in a and its width is essentially given by the temporal convolution of pump and probe pulses. After ~0.2 ps long-lasting changes to O 2p and to a smaller degree also Ni 3d orbitals are visible. The O XAS signal has been measured at 533.8 eV x-ray energy while the Ni XAS was obtained for 851.6 eV (see panels (a) and (c)). 
    \label{fig:XAS}}
 \end{figure}
\begin{figure}
     \includegraphics[width=\linewidth]{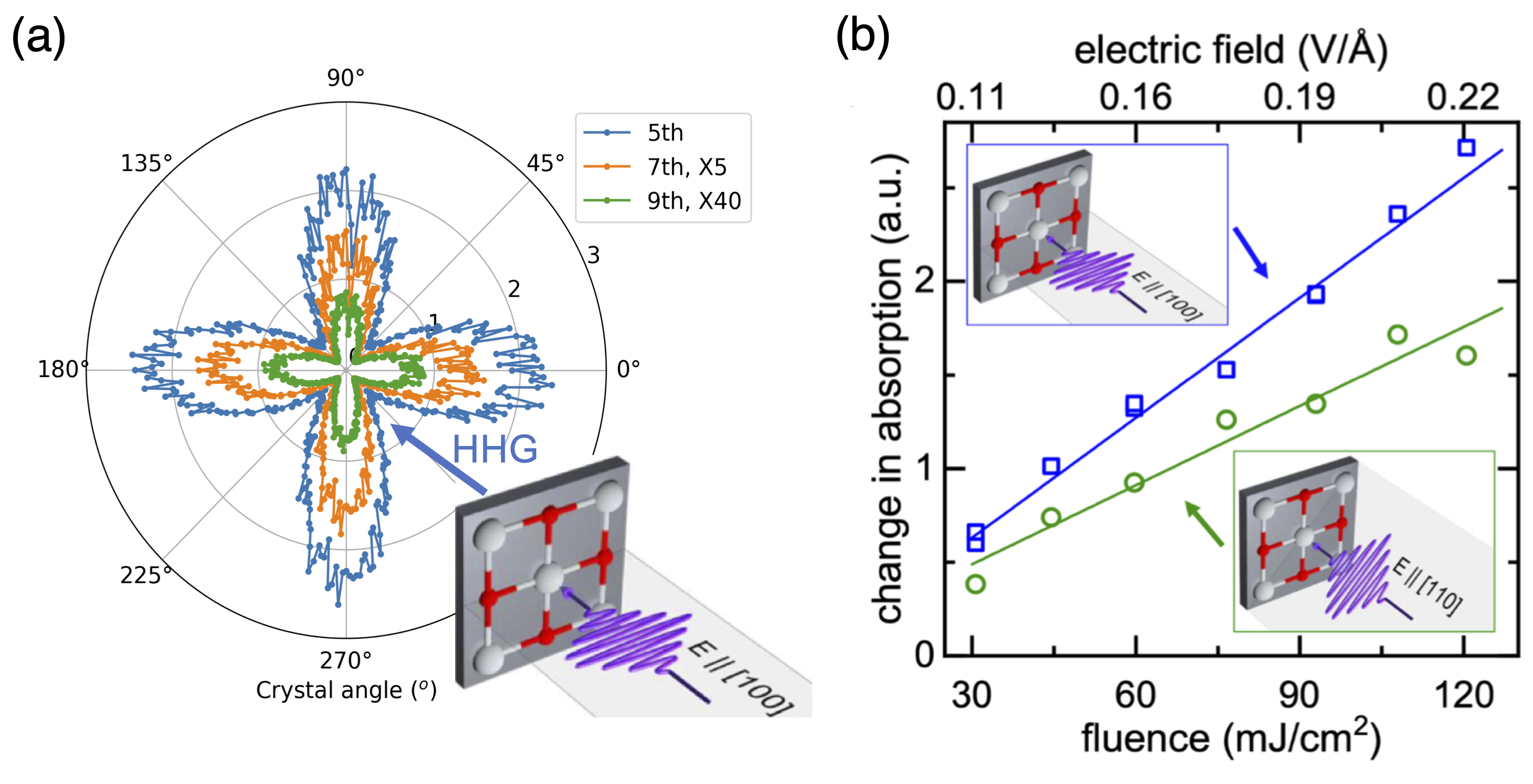}
    \caption{(a) Measurements of the high-harmonics of the driving optical laser field vs. crystal angle relative to the [100] direction as schematically shown in the inset. The measured 5th (blue), 7th (orange) and 9th (green) harmonics show a strong preference for E $||$ [100] directions. The width of the angular distribution decreases with increasing harmonic order. (b) Optical pump – XAS probe measurements as in Fig. \ref{fig:XAS}a for two crystal directions as indicated in the insets. Shown are the pump-laser fluence (bottom axis) and electric peak field (top axis) dependence of the pumped-unpumped difference in O 1s-2p XAS intensity measured at an x-ray energy of 533.8 eV.
    \label{fig:HHG}}
 \end{figure}

We summarize the results of the pump-probe XAS experiment for the O K- and Ni L- edge in Fig. \ref{fig:XAS}. The XAS spectra were taken at the SXR instrument of the LCLS x-ray free-electron laser in Stanford/USA. Soft x-ray probe pulses of 50 fs duration with the x-ray energy tuned to the O 1s-2p (~535 eV photon energy) and Ni 2p-3d (~853 eV) resonances, were incident 20$^{\circ}$ to the normal direction of ultrathin films on NiO grown epitaxially onto single-crystalline MgO(001) substrates (see supplemental material section I.A-C for further details). The driving pump pulses (150 fs duration, 2 $\mu$m central wavelength) were propagating collinearly with the x-rays. The pump polarization was varied during the experiment (see insets of Fig. \ref{fig:HHG}b) while the x-ray polarization was kept fixed along the [100] NiO crystalline direction.   

Figures \ref{fig:XAS}a,b show XAS spectra taken in temporal coincidence with the pump laser pulses. It is evident that there is a laser-induced modification of the O 2p conduction band (Fig. \ref{fig:XAS}a). However, changes of the Ni 3d-states are below the experimental detection limit (Fig. \ref{fig:XAS}b). 
 
The changes visible in the O 1s-2p XAS of Fig. \ref{fig:XAS}a indicate that, when the laser-pump and x-ray-probe pulses arrive simultaneously, a spectral weight transfer takes place in the electric field modified conduction band away from the peaks observed for the ground state. This is clearly visible in the difference spectra (blue line in Fig. \ref{fig:XAS}a) where the XAS peak intensity is attenuated while spectral weight appears at the wings of the peak both at higher and lower x-ray energies. Such a spectral broadening is reminiscent of the driving field cycle resolved dynamical Franz-Keldysh effect observed with attosecond spectroscopy in semiconductors \cite{Lucchini:2016ei}. In the dynamical Franz-Keldysh effect, characteristic changes in the band-structure are expected to be proportional to the square of the electric field, hence with a periodicity of a half-cycle of the pump \cite{Nordstrom:1997jm}. We note that in our experiments we do not resolve the spectral changes during individual cycles of the laser driving field but rather average over the probe-pulse duration, amounting to about 12 half-cycles of the pump within the FWHM of the probe. The temporal evolution of the XAS changes vs pump-probe time-delay at the absorption onset (where the changes are expected to be largest) is depicted in Fig. \ref{fig:XAS}c. Note that despite the signal being the average over many cycles, the intensity changes are significant. We also note that the changes are transient in nature, with close to zero-delay at the experimental time-resolution, and only about 10\% of the transient remains after the pump-pulse. Novelli \emph{et al.} investigates pumping of sub-gap states in the charge-transfer compound La$_2$CuO$_4$, using a 0.95 eV photon energy for a band-gap of 1.8eV. The response, as measured by transient reflection, persists over the 1.5 ps time-scale. A band-gap renormalization persisting after the pump is reported, significantly larger than the changes seen in our measurements of NiO. They attribute this response to interaction with a bosonic field.\cite{Novelli:2013gk}

Electric-fields of sufficient strength may lead to high harmonic generation (HHG). The relation between the polarization direction of a linearly-polarized pulse and the crystal lattice can be used to determine if multi-photon vertical transitions (governed by an anisotropic effective mass tensor) or charge-transfer dynamics (governed by the anisotropy of the real-space orbital structure) dominates the response, similar to what happens in MgO \cite{You:2016hx}. HHG is a complementary measure to the site-local probe that XAS constitutes, and measures the details of electron and hole dynamics in the hybridizing band-structure. In the scenario dominated by the dynamical hopping renormalisation, electrons are accelerated along the direction of optical polarization, where an anisotropic response would indicate an anisotropy in the orbital structure. The polarization of the electronic bands leads to charge-transfer, and a transient broadening of the electronic levels via the dynamical Franz-Keldysh effect \cite{Lucchini:2016ei, Schiffrin:2012kr}. 
This is characterized by the so-called Keldysh parameter ($\sim$5 in our case, see supplemental material II.A). In such conditions the electrons adjust non-adiabatically to the external field, resulting in a distorted band-structure.

To address the effects of HHG generation, we study the crystal orientation dependence of HHG from monocrystalline NiO samples that are subjected to driving optical laser fields (see supplemental material I.D). Results are shown in Fig. \ref{fig:HHG}a. We find that the measurements show maxima along cubic directions, i.e. when the laser field is parallel to [100] direction. In fact, the signal along [100] is about an order of magnitude higher than that along [110]. This behavior is consistent with a microscopic mechanism where a highly preferential charge transfer occurs in the direction of the O 2p - Ni 3d bonds. The direction of the O 2p-states at the valence band maximum is pinned by the hybridization to Ni 3d-states. However, the absence of Ni-3d response in the XAS measurements indicate that the charge-transfer involves final-state orbitals of different character resulting from the non-adiabatic response of the band-structure. As is shown by the transient density response below, time-dependent density functional theory suggests that these states are delocalized, with substantial response in the interstitial region. Because the high-harmonic generation process is extremely non-linear such a pronounced affect can be observed. In Fig. \ref{fig:HHG}a it is seen that the angular width of harmonic decrease significantly with harmonic order, approaching only few degrees for 9th order harmonic. This behavior is reminiscent of that observed in the uncorrelated insulator MgO indicating a more directional charge transfer for higher order harmonics \cite{You:2016hx}. 

Consistent with the angular dependence observed for high harmonic generation we see that the changes in the O 1s-2p XAS are more pronounced when the pump polarization is along the [100] rather than the [110] direction (see Fig. \ref{fig:HHG}b). This demonstrates that it is the O 2p orbitals that display a pronounced directional polarizability in response to the external electric field. A rationale for this is the chemical bonding of NiO, where a direct overlap of Ni 3d- and O 2p-states forms Ni-O-Ni bonds in the [100] direction. The [110] direction is rotated by 45$^{\circ}$, so that the orbital overlap between Ni 3d and O 2p states is weaker. We also note that the change in absorption is linear in fluence, and not electric field strength, indicative of the dynamical Franz-Keldysh effect. Fitting the absorption change to fluence results in exponents of 1.03 and 1.06 for [100] and [110] respectively, leaving little room for effects of any other order, for example changes in long-range screening, linear absorption processes or high-order photon-matter interaction. 

\begin{figure}
     \includegraphics[width=\linewidth]{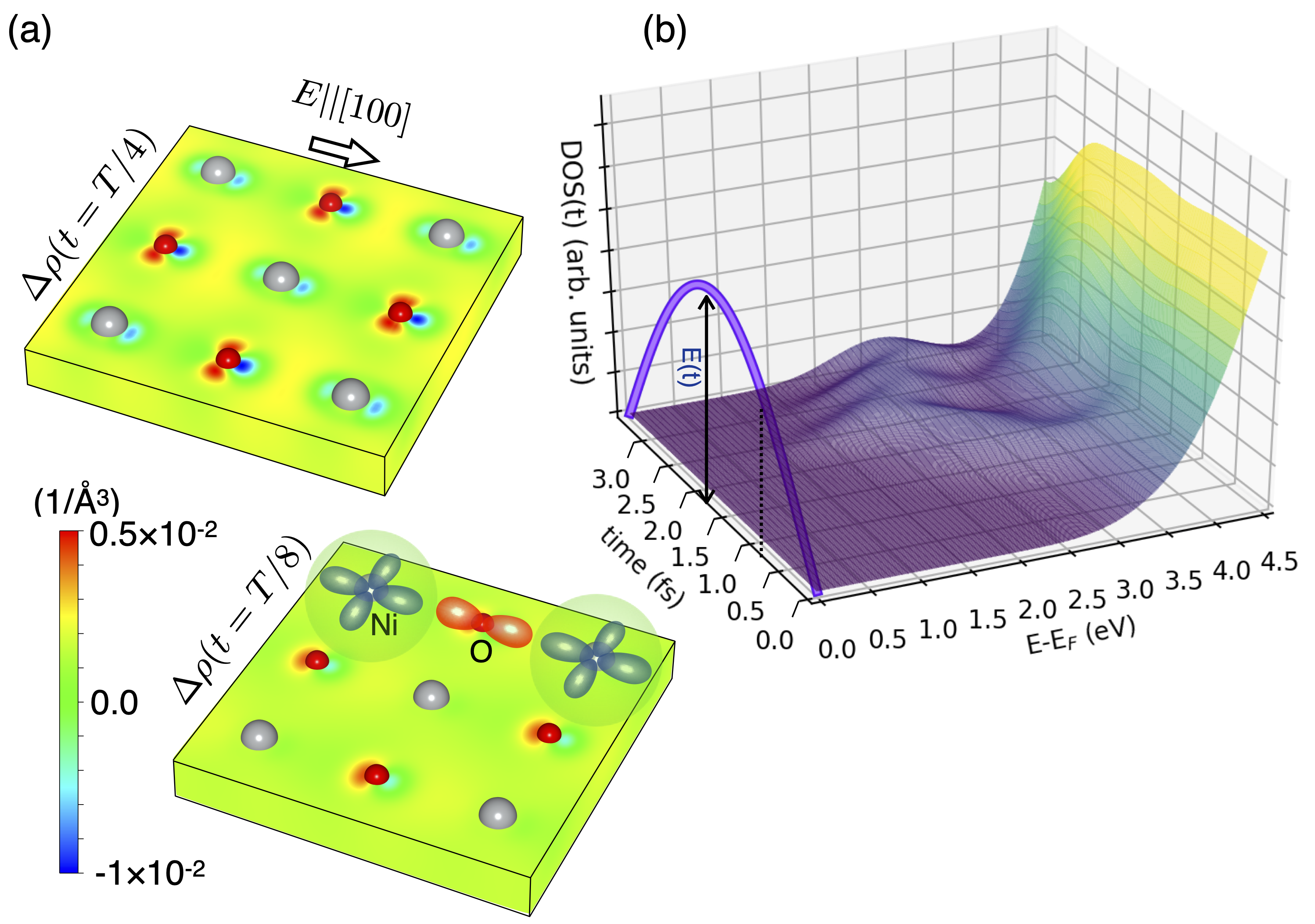}
    \caption{(a) Calculated charge-density difference in real-space as a function of time. The colour represents the difference between the time-dependent and ground-state electron density, with the scale according to the colour-bar in units of electrons/\AA$^3$. Yellow to red indicates an increase in electron density, and cyan to blue a decrease. (b) Calculated change in single-particle energy expectation value close to the conduction band edge, when driven with one half optical cycle of the pump, as indicated by the electric field, E(t), shown as blue line. The colour of the electronic density of states (DOS) indicates the number of states per unit energy. 
    \label{fig:tddft}}
 \end{figure}

The conduction-band dispersion is similar in the two directions  $\Gamma-X$ and $\Gamma-K$ (see Fig. S2). Likewise, the minor difference in band-curvature at the top of the valence band indicates that the effective mass tensor is too symmetric to account for the anisotropy of the harmonic emission in terms of transition rates and we conclude that the asymmetry of the HHG spectrum as a function of polarization must result from field-induced charge transfer, similar to the case of MgO \cite{You:2016hx}. To analyze the physical process in detail, we modelled the dynamical hopping renormalization response of NiO using time-dependent density functional theory, including the external optical field as a vector potential $A_{\text{Ext}}$, (see supplemental material II.C). The fact that the experimentally measured response of the pump-pulse is strong for the O 2p-states, whereas the Ni 3d-states are inert to the pump, allowed us to make the simplifying assumption that U is time-independent. Fig. \ref{fig:tddft} displays the results calculated for one half-cycle of a sine-wave driving field with duration and amplitude identical to the applied external field used in the experimental setup. Initially we focus on changes of the charge density at the Ni and O sites, shown in Fig. \ref{fig:tddft}a by snapshots of the density-difference $\Delta\rho(t)=\rho(0)-\rho(t)$ at T/8 T/4, where T is the period time of 6.9 fs. The NiO charge density response shows that there is a transient transfer of charge, predominantly from the vicinity of the O-atom to the interstitial region, according to the direction of the applied electric field. It is also clear that a far larger response can be seen in the vicinity of O atoms compared to Ni, consistent with the experiments. An understanding of this result can be found in the stronger localization of the Ni 3d states, enforced by the strong attractive Coulomb interaction between the positive Ni nucleus and the nearly localized Ni 3d electrons. The more delocalized oxygen 2p states are dominated by the kinetic energy term, which is directly modified by $A_{\text{Ext}}$, as opposed to the Ni 3d states, that are dominated by the potential term while the kinetic modification has less impact. 

We now relate the calculated changes in electron density (Fig. \ref{fig:tddft}a) to the measured transient XAS (Figs. \ref{fig:XAS},\ref{fig:HHG}) by calculating the time evolution of the electronic density of states (DOS) as detailed in the supplemental material section II.C. Figure \ref{fig:tddft}b shows the DOS for energies close to the conduction band edge at times corresponding to half a cycle of $A_{\text{Ext}}$. At the start of the optical cycle (t = 0 fs) the DOS increases as the conduction band edge is approached above an energy of $E - E_{F} \approx 2.5$ eV. However, as $A_{\text{Ext}}$ increases with time Fig. \ref{fig:tddft}b shows that the band-edge moves to significantly lower energy. There are, as an example, states all the way down to 1.5 eV at t = 2.4 fs. This clearly illustrates a significant band-gap modification, expected from the dynamical Franz-Keldysh effect \cite{Lucchini:2016ei}. In addition, Fig. \ref{fig:tddft}b demonstrates that the appearance of states within the band-gap lags behind the electric driving field. While $A_{\text{Ext}}$ reaches its maximum value at t = 1.7 fs, the in-gap states are most pronounced around 2.4 fs. This 0.7 fs delay can be attributed to the contribution of dielectric polarization currents which counteract $A_{\text{Ext}}$ \cite{Schultze:2012hs, Schiffrin:2012kr, Kruchinin:2018gna}.


Experimental and theoretical results demonstrate that the response of NiO to strong sub-resonant electromagnetic fields is predominantly due to transient modifications to the hopping parameter consistent with a transient directional charge transfer away from O to the interstitial region and Ni. This is observed both as a dynamical Franz-Keldysh effect in the O 1s-2p XAS and a directional strong anisotropy in high harmonic generation experiments, favouring emission when the pump is polarized along the Ni-O-Ni bonding. The absence of change in Ni 2p-3d XAS indicates that the charge-transfer does not alter the electronic structure of 3d-character close to the Ni core region. The agreement between our HHG and XAS resuts indicate that the response of NiO under the current conditions is reminiscent of that of a conventional semicontuctor, such as MgO \cite{You:2016hx}.

The apparent absence of response to the pump of the localized Ni 3d states indicates that the screening-processes responsible for the Hubbard $U$ is not modified on the experimental timescales considered here. This holds for accumulative effects throughout the pump duration of 150 fs, as well as for changes related to the dynamical Franz-Keldysh effect, with the relevant timescale of our pump half-cycle of 3.4fs. Furthermore, the significant modulation of more dispersive states on ultrashort timescale, compared to the inert screening of the Hubbard U, indicates that dispersive states responds much faster, and are more easily influenced by the optical field under the present experimental conditions. The dispersive states are the major contributor to screening of Hubbard U, hence the absence of modulation of Hubbard U must be related to the retardation of screening effects. Thus, a off-resonant driving field has for NiO, and potentially for all correlated electron systems, the largest impact on dispersive, delocalized states. In this regard, the study presented here tests the speed-limit of screening of the on-site Coulomb repulsion.  In the context of cluster models, that are very successful in reproducing the many-body dynamics of the correlated orbitals, these dispersive bands are projected out in the construction of the low-energy Hamiltonian. 

The results presented here indicate that non-local effects accommodating the screening-contribution of the itinerant states are crucial when developing theoretical and computational models for driven dynamics. For NiO the field induced modification of the electronic structure causes a renormalization of the band-gap. The response is strictly transient, (see Fig. \ref{fig:XAS}c), indicating that no energy is pumped in to the system. The absence of response detected for the Ni L-edge indicates that although itinerant states polarize, this does not propagate to the screening of the Ni 3d-states. This finding has significant implications for future experimental and theoretical developments, and potentially for technological applications.
In light of the Novelli \emph{et al.} \cite{Novelli:2013gk}, the strictly transient response in NiO indicates that there are no bosonic degrees of freedom that responds to off-resonant fields on similar time-scales as those of La$_{2}$CuO$_{4}$.

Since the Ni-O-Ni bond is responsible for mediating the super-exchange interaction that governs the equilibrium antiferromagnetic order in NiO \cite{Kramers:1934bf, Anderson:1950wk, Goodenough:1955dm}, any transient modification of the density around the oxygen atom that breaks the Ni-O-Ni symmetry, as implied by the inset of Fig. \ref{fig:tddft}a, must change this magnetic interaction in the resulting non-equilibrium state. It is possible that the modification of the exchange interaction will set the magnetic configuration in motion as dictated by the Landau-Lifshitz-Gilbert equation. Modifications of the magnetic exchange by sub-gap excitation have been observed using all-optical pump-probe methods \cite{Kirilyuk:2010ha}. Here we provide the x-ray analogue that clearly demonstrates how the electronic structure is modified by a directional transient charge transfer along the O-Ni bond. These results, in addition to the delayed intra-cycle response predicted by theory, point to the importance of atto-second probes to resolve the response to the pump on intra-cycle timescales. Future measurements will be able to probe what type of possible magnetic excitation evolve on ultrafast timescales \cite{Iacocca:2019hh}. The theoretical approach described here has potential to lead the way to guide and interpret such measurements.

\begin{acknowledgments}
OG acknowledge financial support from the Strategic Research Council (SSF) grant ICA16-0037 and the Swedish Research Council (VR) grant 2019-03901. The computations were enabled by resources provided by the Swedish National Infrastructure for Computing (SNIC), partially funded by the Swedish Research Council through grant agreement no. 2018-05973. OE acknowledges support from the Swedish Research Council (VR), the Knut and Alice Wallenberg (KAW) foundation, the Foundation for Strategic Research (SSF), the European Research Council (854843-FASTCORR ), and eSSENCE. HAD acknowledges support from the Swedish Research Council (VR) grants 2017-06711and 2018-04918. MB and RYE are funded by the Helmholtz Association through grant VH-NG-1105. IV acknowledges support by the U.S. Department of Energy, Office of Science, Office of Basic Energy Sciences under the X-Ray Scattering Program Award Number DE-SC0017643. Operation of LCLS is supported by the U.S. Department of Energy, Office of Basic Energy Sciences under contract No. DE-AC02-76SF00515. Work at Pulse was supported by the AMOS program within the Chemical Sciences Division of the Office of Basic Energy Sciences, Office of Science, U.S. Department of Energy. SG, JL and DAR acknowledge the support and SG additionally acknowledges the Early Career Research Program. Work at CFEL was supported by the European Research Council (ERC-2015-AdG-694097), Grupos Consolidados (IT1249-19) and the Flatiron Institute, a division of the Simons Foundation. We acknowledge funding by the Deutsche Forschungsgemeinschaft (DFG) under Germany’s Excellence Strategy - Cluster of Excellence Advanced Imaging of Matter (AIM) EXC 2056 - 390715994 and funding by the Deutsche Forschungsgemeinschaft (DFG). MAS acknowledges funding through the Emmy Noether Programme of Deutsche Forschungsgemeinschaft (SE 2558/2-1).
\end{acknowledgments}

\bibliography{NiO_LCLS}

\end{document}